\documentclass[conference]{IEEEtran}
\usepackage{times,amssymb,amsmath,amsfonts,float,nicefrac,color}
\usepackage{euscript,graphics}
\usepackage[dvips]{epsfig}

\newtheorem{theorem}{Theorem}[section]

\newtheorem{lemma}[theorem]{Lemma}

\newtheorem{cnstr}{Construction}

\newtheorem{xmpl}{Example}

\newcommand\qed{\rule{1mm}{2mm}\medskip}

\newcommand{\remove}[1]{}

\newcommand\nd{\noindent}%\newcommand\bysame{\rule[1mm]{1cm}{.025cm}}

\newcommand{\ceilenv}[1]{\left\lceil #1 \right\rceil}

\newcommand\nc\newcommand
\nc\bfa{{\boldsymbol a}}\nc\bfA{{\bf A}}\nc\cA{{\mathcal A}}
\nc\bfb{{\boldsymbol b}}\nc\bfB{{\bf B}}\nc\cB{{\mathcal B}}
\nc\bfc{{\boldsymbol c}}\nc\bfC{{\bf C}}\nc\cC{{\mathcal C}}
\nc\bfd{{\boldsymbol d}}\nc\bfD{{\bf D}}\nc\cD{{\mathcal D}}
\nc\bfe{{\boldsymbol e}}\nc\bfE{{\bf E}}\nc\cE{{\mathcal E}}
\nc\bff{{\boldsymbol f}}\nc\bfF{{\bf F}}\nc\cF{{\mathcal F}}
\nc\bfg{{\boldsymbol g}}\nc\bfG{{\bf G}}\nc\cG{{\mathcal G}}
\nc\bfh{{\boldsymbol h}}\nc\bfH{{\bf H}}\nc\cH{{\mathcal H}}
\nc\bfi{{\boldsymbol i}}\nc\bfI{{\bf I}}\nc\cI{{\mathcal I}}
\nc\bfj{{\boldsymbol j}}\nc\bfJ{{\bf J}}\nc\cJ{{\mathcal J}}
\nc\bfk{{\boldsymbol k}}\nc\bfK{{\bf K}}\nc\cK{{\mathcal K}}
\nc\bfl{{\boldsymbol l}}\nc\bfL{{\bf L}}\nc\cL{{\mathcal L}}
\nc\bfm{{\boldsymbol m}}\nc\bfM{{\bf M}}\nc\cM{{\mathcal M}}
\nc\bfn{{\boldsymbol n}}\nc\bfN{{\bf N}}\nc\cN{{\mathcal N}}
\nc\bfo{{\boldsymbol o}}\nc\bfO{{\bf O}}\nc\cO{{\mathcal O}}
\nc\bfp{{\boldsymbol p}}\nc\bfP{{\bf P}}\nc\cP{{\mathcal P}}
\nc\bfq{{\boldsymbol q}}\nc\bfQ{{\bf Q}}\nc\cQ{{\mathcal Q}}
\nc\bfr{{\boldsymbol r}}\nc\bfR{{\bf R}}\nc\cR{{\mathcal R}}
\nc\bfs{{\boldsymbol s}}\nc\bfS{{\bf S}}\nc\cS{{\mathcal S}}
\nc\bft{{\boldsymbol t}}\nc\bfT{{\bf T}}\nc\cT{{\mathcal T}}
\nc\bfu{{\boldsymbol u}}\nc\bfU{{\bf U}}\nc\cU{{\mathcal U}}
\nc\bfv{{\boldsymbol v}}\nc\bfV{{\bf V}}\nc\cV{{\mathcal V}}
\nc\bfw{{\boldsymbol w}}\nc\bfW{{\bf W}}\nc\cW{{\mathcal W}}
\nc\bfx{{\boldsymbol x}}\nc\bfX{{\bf X}}\nc\cX{{\mathcal X}}
\nc\bfy{{\boldsymbol y}}\nc\bfY{{\bf Y}}\nc\cY{{\mathcal Y}}
\nc\bfz{{\boldsymbol z}}\nc\bfZ{{\bf Z}}\nc\cZ{{\mathcal Z}}
\nc\od{{\bar d}}\nc\ow{{\bar w}}\nc\odelta{{\bar\delta}}
\nc\ox{{\bar x}}\nc\oy{{\bar y}}\nc\ou{{\bar u}}
\nc\oh{{\bar h}}
\DeclareMathOperator{\E}{\text{\sf E}} 
\DeclareMathOperator{\C}{Cl}

\newcommand\ff{{\mathbb F}}

\nc\ellone{{\ell_1}}
\nc\elltwo{{\ell_2}}
\nc\ellinf{{{\ell_\infty}}}
\nc\ip[2]{\langle #1,#2\rangle}

\newcommand{\beeq}{\begin{eqnarray*}}
\newcommand{\eneq}{\end{eqnarray*}}

\begin{document}

\title{Bounds on Locally Recoverable Codes with Multiple Recovering Sets} 

\author{\IEEEauthorblockN{Itzhak Tamo$^\ast$}
\and \IEEEauthorblockN{Alexander Barg$^{\ast}$}}
\maketitle
%\vspace*{-.5in}
{\renewcommand{\thefootnote}{}\footnotetext{

\vspace{-.2in}
 
\noindent\rule{1.5in}{.4pt}

\nd$^\ast$ Department of ECE and Institute for Systems Research, University of Maryland, College Park, MD 20742. Emails: 
\{zactamo,alexanderbarg\}@gmail.com. Research supported by NSF grants CCF1217894, CCF1217245, and by NSA grant 98230-12-1-0260.
}}
\renewcommand{\thefootnote}{\arabic{footnote}}
\setcounter{footnote}{0}

%\vspace*{-.5in}
\begin{abstract}
A locally recoverable code (LRC code) is a code  over a finite alphabet such that every symbol in the encoding is 
a function of a small number of other symbols that form a recovering set.
Bounds on the rate and distance of such codes have been extensively studied in the literature. In this paper we derive
upper bounds on the rate and distance of codes in which every symbol has $t\ge 1$ disjoint recovering sets.

\end{abstract}

\section{Introduction}
Locally recoverable (LRC) codes currently form one of the rapidly developing topics in coding theory because of their applications in
distributed and cloud storage systems. Recently LRC codes have been the subject of a large number of publications, among them
\cite{Dim2000,Gop11,Pap12,Ras11,Sil11,tam13,Gop13}. 
We say that a code $\cC\subset \ff_q^n$ has locality $r$ if every symbol of the codeword $x\in \cC$ 
can be recovered from a subset of $r$ other symbols of $x$ (i.e., is a function of some other $r$ symbols $x_{i_1},x_{i_2},\dots,x_{i_r}$).
In other words, this means that, given $x\in \cC, i\in [n],$ there exists a subset of coordinates 
$R_i\subset [n]\backslash i, |R_i|\le r$ such that
the restriction of $\cC$ to the coordinates in $R_i$ enables one to find the value of $x_i.$ The subset 
$R_i$ is called a {\em recovering set} for the symbol $x_i$. 
%We denote the  $\cC(n,k,r)$ \textcolor{red}{(I don't like this notation $\cC(n,k,r)$), I prefer $(n,k,r)$ LRC. What about the notation $(n,k,r,t)$ LRC?)} to denote a code with locality $r$ of length $n$ and cardinality $q^k$ over the alphabet of $q$ symbols.

Now assume that every symbol of the code $\cC$ can be recovered from $t$ disjoint subsets of symbols of size $r_1,\dots,r_t$ respectively,
called recovering sets of the symbol.
Below we shall restrict ourselves to the case $r_1=\dots=r_t=r$ which makes the expressions of the bounds more compact. At the same time,
we note that the technique presented below enables us to treat the general case as well.
Given a code $\cC$ with $t$ disjoint recovering sets of size $r$, we use the notation 
$(n,k,r,t)$ to refer to its parameters. If the values of $n,k,r$ are understood, we simply call $\cC$ a $t$-LRC code.

More formally, denote by $\cC_I$ the restriction of the code $\cC$ to a subset of coordinates $I\subset [n].$
Given $a\in \ff_q$ define the set of codewords
   $
   \cC(i,a)=\{x\in \cC: x_i=a\},\; i\in[n].
   $

\vspace*{.1in}
\nd{\sc Definition:} A code $\cC$  is said to have $t$ disjoint recovering sets if for every $i\in[n]$ there are $t$ pairwise disjoint subsets $R_{i,1},\dots,R_{i,t}\subset [n]$ such that for all $j=1,\dots,t$
  $$
  \cC_{R_{i,j}}(i,a)\cap \cC_{R_{i,j}}(i,a')=\emptyset,\quad  a\ne a'.
  $$
%  $\cC(n,k,\{r_1,r_2,\dots,r_t\})$ code
 
 Having more then one recovering set is beneficial in practice because it enables more users to access a given portion of data, thus enhancing data availability in the system.

%If the code has two recovering sets for each symbol, we use the notation $\cC(n,k,\{r,r\}).$

One of the main questions studied for LRC codes is related to estimates of
the largest possible minimum distance of codes with locality $r$. 
\begin{theorem} \label{thm:S}
Let $\cC$ be an $(n,k,r,t=1)$ LRC code, then: \\
   The rate of $\cC$ satisfies
				\begin{equation}
							\frac{k}{n}\leq  \frac{r}{r+1}.
							\label{prop546}
				\end{equation}
				The minimum distance of $\cC$ 
				satisfies 
   			\begin{equation}
				d\leq n-k-\ceilenv{ \frac{k}{r}}+2.				 
				\label{eq:erer}
     		\end{equation}
\end{theorem}
These upper bounds on the distance and rate of LRC codes were proved in \cite{Gop11,Pap12}.
Recently codes that generalize Reed-Solomon codes and achieve the bound \eqref{eq:erer} for any $n$ were constructed in \cite{tam13}.
Other bounds on the distance of LRC codes appear in  \cite{Cad13,pra14}.

A graph-theoretic proof of Theorem \ref{thm:S} was recently presented in \cite{tam13}. Developing the ideas of this paper, here we prove the following results.
\begin{theorem} \label{thm:2s}
Let $\cC$ be an $(n,k,r,t)$ LRC code with $t$ disjoint recovering sets of size $r$. Then the rate of $\cC$ satisfies
\begin{equation}
\frac{k}{n}\leq \frac{1}{\prod_{j=1}^t(1+\frac{1}{jr})}.
\label{cvcv}
\end{equation}
The minimum distance of $\cC$ is bounded above as follows:
   \begin{equation}\label{eq:d2}
  d\leq n-\sum_{i=0}^t\Big\lfloor\frac{k-1}{r^i}\Big\rfloor.
  \end{equation}
\end{theorem}
{\sc Remarks:} \\1. For $t=1$ the bound on the rate \eqref{cvcv} reduces to \eqref{prop546}. For general $t$ the expression 
on the right-hand side of \eqref{cvcv} approximately equals $t^{-1/r}$ (more precise results
 are established below in the paper; see Lemma \ref{lemma:rroot}). 

\vspace*{.05in}\nd 2. {\sc On tightness of the bound on the rate} \eqref{cvcv}. For a code with a single recovering set for every symbol, inequality \eqref{cvcv} provides a tight bound on
the rate \eqref{prop546}. 
For two recovering sets the bound \eqref{cvcv} takes the form 
   \begin{equation}\label{eq:r2r}
   \frac{k}{n}\leq \frac{2r^2}{(r+1)(2r+1)}.
   \end{equation}
Addressing the question of the tightness of the bound consider a binary code which is the product of
two single-parity-check codes with ${r}$ message symbols each. The resulting rate equals $r^2/(r+1)^2$ which is only
slightly less than the right-hand side of \eqref{eq:r2r}. 

Generalizing, we can construct a $t$-fold power of the binary $(r+1,r)$ single-parity-check code and obtain a code with $t$ disjoint 
recovering sets that has the rate $(r/(r+1))^t.$ We believe that the rate $(r/(r+1))^t$ is the largest possible for a code with
$t$ disjoint recovering sets as long as $t$ is not too large (e.g., $O(\log n)$).
%\textcolor{red}{(we need to say that it is true for $t$ not too big.)}. 

\remove{
For any $r^2$ information symbols $a_{i,j},i,j\in [r]$ we add extra $2r+1$ parity symbols as follows: Let $A=(a_{i,j})_{i,j\in [r+1]}$ be an $(r+1)\times (r+1)$ matrix whose entries in the $(r+1)$-th row and column satisfy that the entries in each row and column sum up to zero. It is clear that for any $r^2$ information symbols, $(r+1)^2$ symbols are stored, hence the rate is as claimed. Moreover, each symbol can be recovered in two disjoint ways (either by the remaining symbols of its row or column). This simple construction can be generalized to any number of recovering sets $t$, with rate $r^t/(r+1)^t$. We believe that this rate is the best possible, namely this is a tight bound.}

\vspace*{.05in}\nd 3. {\sc On tightness of the bound on the distance} \eqref{eq:d2}. For $t=1$ the bound \eqref{eq:d2} reduces to \eqref{eq:erer}, and there exist large families of codes that meet this bound with equality \cite{tam13,Sil11,My-paper}.
The next interesting case, in particular for applications, is $t=2.$  From \eqref{eq:d2} we obtain the bound
  \begin{equation}\label{2}
  d\leq n-\Big( k-1+\Big \lfloor \frac{k-1}{r}\Big\rfloor+\Big\lfloor \frac{k-1}{r^2}\Big\rfloor\Big).
  \end{equation}
Interestingly, this bound is also tight. Indeed, consider the shortened binary Hamming code of length $6$ with
the parity-check matrix

 {\small $$
  \begin{pmatrix}
  0&0&0&1&1&1\\
  0&1&1&0&0&1\\
  1&0&1&0&1&0
  \end{pmatrix}.
  $$}
\noindent It is easily seen that this is a $(6,3,\{2,2\})$ LRC code, and its distance $d=3$ meets the bound \eqref{2} with equality.

\section{An Upper Bound on the Rate of LRC Codes}
%\section{Proof of Theorem \ref{thm:2s}}
\subsection{The recovering graph}
\label{the recovering graph}
Assume that coordinate $i$ has $t$ disjoint recovering sets $R^i_1,...R^i_t$,  each of size $r$, where $R^i_j\subset [n]\backslash i$.
Define a directed graph $G$ as follows. The set of vertices $V=[n]$ corresponds to the set of $n$ coordinates of the LRC code. The ordered pair of vertices $(i,j)$ forms a directed edge $i\to j$ if $j\in R^i_l$ for some $l\in[t].$ 
We color the edges of the graph with $t$ distinct colors in order to differentiate between the recovering sets of each coordinate. 
More precisely, let $F_e:E(G)\to[t]$ be a coloring function of the edges,
given by $F((i,j))=l$ if $j\in R^i_l.$ 
%i.e., if $j$ is contained in the $l$-th recovering set of the vertex $i$, where $1\leq l \leq t$. 
Thus, the out-degree of each vertex $i\in V=V(G)$ is $\sum_{l}|R^i_l|=tr,$ and the edges leaving $i$ are colored in 
$t$ colors. We call $G$ the {\em recovering graph} of the code $\cC.$

The following lemma will be used to prove the main theorem of this section.
\begin{lemma} \label{lemma:rfrf} 
There exists a subset of vertices $U\subseteq V$ of size at least 
  \begin{equation}\label{eq:size}
  |U|\geq n\Big(1-\frac{1}{\prod_{j=1}^t(1+\frac{1}{jr})}\Big)
  \end{equation}
such that for any $U'\subseteq U$, the induced subgraph $G_{U'}$ on the vertices $U'$ 
has at least one vertex $v\in U'$ such that its set of outgoing edges $\{(v,j),j\in U')\}$ is missing at least one color. 
%There exists a permutation $\tau:V\to V$ and a subset $U\subset V$ of cardinality  
%such that the induced subgraph of every subset $U'\subseteq U$ contains a vertex such that the number of colors of its outgoing edges is at most $t-1$.
\end{lemma}

\begin{IEEEproof}
For a given permutation $\tau$ of 
the set of vertices $V=[n]$, we define the coloring of some of the vertices as follows:  
The color $j\in [t]$ is assigned to the vertex $v$ if 
%the value $v$ attains in $\tau$ is greater than the values its $j$-th recovering sets attains  in $\tau$, i.e., 
   \begin{equation}
     \tau(v)>\tau(m) \quad\text{for all }m\in R^v_j.
   \label{eq:dfdf}
   \end{equation}
If this condition is satisfied for several recovering sets $R^v_j$, the vertex $v$ is assigned any of the colors $j$ corresponding to these sets. 
Finally, if this condition is not satisfied at all, then the vertex $v$ is not colored.

Let $U$ be the set of colored vertices, and consider one of its subsets $U'\subseteq U$.
\remove{For a color $i,1\le i\le t$ and a vertex $v$ denote by 
$N_i(v)$ the set of vertices $u$ such that $F_e((v,u))=i,$ i.e., $u\in R_i^v.$

Let $\tau$ be some permutation on $n$ elements and let $U$ be the set of all the vertices $v\in V$ such that there exists $i\in[t]$ with
the property that $\tau(v)$ is greater than
all the values $\tau(u), u\in N_i(v),$ i.e., 
  \begin{equation}\label{eq:ti}
   \tau(v)> \max_{u\in N_i(v)}\tau(u).
  \end{equation}
Pick a random permutation $\tau$ for the set of vertices $V$, and define the set $U$ to contain all the vertices $v$ such that $\tau(v)$ is greater than the values given by $\tau$ to at least one color set $N_i(v),$ namely there exists 
$i$ such that 
%$>>>>>>>>>>>$
Next, color the vertices of $U$ using the $m$ colors in the following manner: A vertex $v$ in $U$ will be colored in color $i$ if 
\begin{equation}
\tau(v)>\max_{u\in N_i(v)}\tau(u).
\label{eq:rdrd}
\end{equation}
Note that if (\ref{eq:rdrd}) is satisfied for several colors, then color the vertex $v$ arbitrarily with one of these colors. By the definition of the set $U$, it is clear that this defines a coloring for all the vertices of $U$. We claim that the set $U$ is the desired set.} 
Let $G_{U'}$ be the induced subgraph on $U'$. We claim that there exists $v\in U'$ such that its set of outgoing edges is missing at least one color in $G_{U'}$. Assume toward a contradiction that every vertex 
of $G_{U'}$ has outgoing edges of all $t$ colors. 
%Define the following path on the graph $G_{U'}$. 
Choose a vertex $v\in U'$ and construct a walk through 
the vertices of $G_{U'}$ according to the following rule. 
If the path constructed so far ends at some a vertex with color $j,$ choose one of its outgoing edges also
 colored in $j$ and leave the vertex moving along this edge. 
 By assumption, every vertex has outgoing edges of all $t$ colors, so this process, and hence this path can be extended indefinitely. 
Since the graph $G_{U'}$ is finite, there will be a vertex, call it $v_1,$ that is encountered twice. 
The segment of the path that begins at $v_1$ and returns to it has the form
   $$
   v_1\rightarrow v_2\rightarrow... \rightarrow v_l,
   $$
where $v_1=v_l$. For any $i=1,...,l-1$ the vertex $v_i$ and the edge $(v_i,v_{i+1})$ are colored with 
the same color. Hence by the definition of the set $U$ we conclude that 
$\tau(v_i)>\tau(v_{i+1})$ for all $i=1,\dots, l-1,$ a contradiction. 

In order to show that there exists such a set $U$ of large cardinality, we choose the permutation
$\tau$ randomly and uniformly among all the $n!$ possibilities and compute the expected cardinality 
of the set $U$
%$>>>>>>>>$
%Let us estimate the expected cardinality of the set $U$ if the permutation $\tau$ is chosen randomly and uniformly. 

Let $A_{v,j}$ be the event that \eqref{eq:dfdf} holds for the vertex $v$ and the color $j.$ Since $\Pr(A_{v,j})$ does not depend
on $v$, we suppress the subscript $v$, and write
  $$
  \Pr(v\in U)=\Pr( \cup_{i=j}^t A_{j}).
  $$
Let us compute the probability of the event $\cup_{j=1}^tA_j.$
Note that for any set $S\subseteq [t]$ the probability of the event that all the $A_j,j\in S$ occur simultaneously, equals 
$$P(\cap_{j\in S}A_j)=\frac{1}{|S|r+1},$$
Hence by the inclusion exclusion formula we get 
\begin{align}
\Pr(\cup_{j=1}^t A_j)&=\sum_{j=1}^t(-1)^{j-1}\binom{t}{j}P(A_1\cap...\cap A_j)\nonumber\\
&=\sum_{j=1}^t(-1)^{j-1}\binom{t}{j}\frac{1}{jr+1}\nonumber\\
&=\frac{-1}{r}\Big(\sum_{j=0}^t(-1)^j \binom{t}{j}\frac{1}{j+\frac{1}{r}}-r\Big)\nonumber\\
&=1-\frac{1}{r}\sum_{j=0}^t(-1)^j \binom{t}{j}\frac{1}{j+\frac{1}{r}}\nonumber\\
&=1-\frac{1}{r}\frac{t!}{\frac{1}{r}(1+\frac{1}{r})...(t+\frac{1}{r})}\label{titi}\\
&=1-\frac{1}{\prod_{j=1}^t(1+\frac{1}{jr})}\nonumber,
\end{align}
where \eqref{titi} follows from \cite[p.~188]{concrete_math}.
Now let $X_v$ be the indicator random variable for the event that $v\in U$, then
\begin{align*}
\E(|U|)&=\sum_{v\in V}\E(X_v)\\
&=\sum_{v\in V}\Pr(v\in U)
%\\
\end{align*}\begin{align*}
&=n\Pr( \cup_{j=1}^t A_{j})\\
%\end{align*}\begin{align*}
&=n(1-\frac{1}{\prod_{j=1}^t(1+\frac{1}{jr})}).
\end{align*}
The proof is completed by observing that there exists at least one choice of $\tau$ for which $|U|\geq \E(|U|).$
\end{IEEEproof}
%Now we are ready to prove the bound on the rate of an LRC code.
\remove{\begin{theorem}
\label{thm1}
The rate of an $(n,k)$ LRC code with $t$ disjoint recovering sets, each of size at most $r$, satisfies
\begin{equation}
\frac{k}{n}\leq \frac{1}{\prod_{j=1}^t(1+\frac{1}{jr})}.
\label{cvcv}
\end{equation}}

\subsection{ Proof of the bound on the rate \eqref{cvcv}}
Let $U\subseteq [n]$ be the set of vertices of cardinality as in \eqref{eq:size} constructed in Lemma \ref{lemma:rfrf} and let $\overline{U}=[n]\backslash U$ be its complement in $[n].$ 
We claim that the value of every coordinate $i\in U$ can be recovered by accessing the coordinates 
in $\overline{U}.$ To show this, we construct the following iterative procedure, which in each step is applied to the subset $U'\subseteq U$
formed of the coordinates whose values are still unknown.
In the first step $U'=U.$ By Lemma \ref{lemma:rfrf} the induced subgraph $G_{U'}$ contains a vertex $v\in U'$ that is missing one color, call it $i$. This means that the $i$-th recovering set of $v$
is entirely contained in $\overline{U'}$. Hence one can recover the value of the coordinate $v$ of the codeword
by 
knowing the values of the coordinates in $\overline{U'}$. In the next step use the same argument for the set 
of coordinates $U'\backslash\{v\}.$ In this way all the coordinates in $U$ are recovered step by step relying only on the values
of the coordinates in $\overline U.$ Therefore, 
    $$
    k\leq |\overline{U}|\leq \frac{n}{\prod_{j=1}^t(1+\frac{1}{jr})}
    $$
and the proof of \eqref{cvcv} is complete.  
 \hfill\qed

To get a clearer impression of the bound on the rate derived, observe that 
$$
\log \prod_{j=1}^t\Big(1+\frac1{jr}\Big)=\sum_{j=1}^t\log\Big(1+\frac1{jr}\Big)\approx \sum_{j=1}^t\frac1{jr}\approx\frac1r \log t.
$$
Therefore, the value of the product in \eqref{cvcv} is about $\sqrt[r]t.$ More precisely, we have
\begin{lemma}\label{lemma:rroot}
$$\sqrt[r]{t+1}\leq \prod_{j=1}^t\Big(1+\frac{1}{jr}\Big)\leq\sqrt[r]{t+1}\Big(1+\frac{1}{r}\Big).$$
Therefore the rate of a $t$-LRC code \eqref{cvcv} satisfies
   \begin{equation*}\label{eq:rroot}
   \frac{k}{n}\leq  \frac{1}{\sqrt[r]{t+1}}
   %1-\frac{1}{\sqrt[r]{t+1}}\leq P(\cup_{i=1}^tA_i)\leq 1-\frac{r}{(r+1)\sqrt[r]{t+1}}.
   \end{equation*}
\end{lemma}

\begin{IEEEproof}
For $i=0,...,r-1$ define the quantity 
  $$
  f_i=\prod_{j=1}^t\Big(1+\frac{1}{i+jr}\Big).
  $$ 
It can be easily seen that for any $i$, 
   \begin{align}
     f_i\leq f_0 &\leq f_i\Big(1+\frac{1}{r}\Big)\Big(1+\frac{1}{(t+1)r}\Big)^{-1} \label{eq:1nnh} \\ \nonumber
     &=f_i\Big(1+\frac{t}{(t+1)r+1}\Big).\nonumber
\end{align} 
Hence
\begin{align}
\prod_{i=0}^{r-1}f_i&=\prod_{i=0}^{r-1}\prod_{j=1}^t(1+\frac{1}{i+jr})\nonumber\\
&=\prod_{j=r}^{(t+1)r-1}(1+\frac{1}{j})\nonumber\\
&=t+1\label{eq:2nnh}.
\end{align}

Using the inequalities \eqref{eq:1nnh} in \eqref{eq:2nnh}, we obtain 
\begin{align*}
\sqrt[r]{t+1}&=\sqrt[r]{\prod_{i=0}^{r-1}f_i}
\leq \sqrt[r]{\prod_{i=0}^{r-1}f_0}\\
&=\prod_{j=1}^t\Big(1+\frac{1}{jr}\Big)\\
%\end{align*}\begin{align*}
&\leq \sqrt[r]{\prod_{i=0}^{r-1}f_i(1+\frac{t}{(t+1)r+1})}\\
&=\sqrt[r]{t+1}(1+\frac{t}{(t+1)r+1})\\
&\leq \sqrt[r]{t+1}(1+\frac{1}{r}).
\end{align*}
\end{IEEEproof}

%has two disjoint recovering sets, simply eqattained On the other hand, an LRC with two recovering sets and rate 

\section{An Upper Bound on the Minimum Distance of LRC Codes: Proof of \eqref{eq:d2} }
 Consider the recovering graph $G$ of an $(n,k,r,t)$ LRC code $\cC$ with $t$ recovering sets,
 defined in Sect.~\ref{the recovering graph}.
 Define the following coloring procedure of the vertices. Start with an arbitrary subset of vertices
 $S\subseteq V$ and color it in some fixed color, call it red. Now let us color some of the remaining uncolored
 vertices according to the following rule. A vertex is colored red if at least one of its recovering sets is completely colored in red.
This process continues until no more vertices can be colored (recall that $G$ is finite).
%Since the graph is finite, \textcolor{green}{this process will eventually terminate} \textcolor{red}{(which process? we didn't say that we repeat the last step as long as we can.)}. 
Call the set of red-colored vertices obtained at this point the {\em closure} of the set $S$ and call the quantity
$|\C(S)|/|S|$ {\em expansion ratio} of the set $S.$
Since the expansion ratio equals to the quotient of the number of coordinates whose value is determined 
by the set $S$ and the size of the set $S$ itself, it is clear that 
large expansion ratio means that the set $S$ contains a large amount of information about the other coordinates 
of the code. In other words, a large number of values of coordinates outside $S$ is determined by the
values of the coordinates in $S.$

Recall the definition of the distance of the code $\cC$ of length $n$ and cardinality $q^k$ over 
an alphabet of size $q$: 
    $$
    d=n-\max_{I\subseteq [n]}\{|I|:|\cC_I|<q^k\},
    $$
where $\cC_I$ is the restriction of the code to coordinates in $I$. Using the recovering graph and 
the expansion ratio concept, we will show that there exists a  large set $I\subseteq [n]$ 
of coordinates such that $|\cC_I|<q^k$. 

We need the following two lemmas whose proofs are deferred to the end of the section.

%\begin{IEEEproof}
%Let $G$ be the recovering graph of the code. Let the colors of the eges be either white or black. Let $G_w$ be the subgraph of $G$ restricted to its white edges. Similarly, define the graph $G_b$.
%Consdiner the graph $G_w$, then by Lemma \ref{lemma:rfrf} with $m=1$ colors, we get that there exists a subset $U\subseteq [n]$ of vertices of size at least $\frac{n}{r+1}$ such that the induced graph of $G_w$ on $U$ is an acyclic graph. Remove from $U$ arbitrary set of vertices, till it will contain exactly $\lfloor \frac{k-1}{r^2} \rfloor$ vertices. Next, define the set $N$ to contain all the vertices in $[n]\backslash U$ that have a least one incoming edge from a vertex in $U$, namely
%$$N=\{v\in [n]\backslash U: v\in N_w(u) \text{ for some } u\in U\}.$$
%Clearly the size of $N$ is at most $|N|\leq |U|r=\lfloor \frac{k-1}{r^2}\rfloor r\leq \lfloor \frac{k-1}{r}\rfloor.$ Note that since $G_{w,U}$ contains no cycles, it is clearly that knowing the value of the coordinates in $N$ gives exactly the value of each coordinate in $U$.    
%
%\end{IEEEproof}

\begin{lemma}
\label{best lemma}
Let $G$ be the recovering graph of a $(n,k,r,t)$ LRC code $\cC$. For any vertex $v\in G$ there exists a set $S$ of size at most $r^t$ such that $v\in \C(S),$ 
and the expansion ratio of $S$ is at least 
  \begin{equation}
e_t=\frac{r^{t+1}-1}{r^{t+1}-r^t}.
    \label{dikla6}
    \end{equation}
\end{lemma}

\begin{lemma}
\label{best lemma2}
Let $m$ be an integer whose base-$r$ representation is
   $$
    m=\sum_{i}\alpha_ir^i,
    $$
then for an integer $t$,
   $$\Big\lfloor \frac{m}{r^t}\Big\rfloor r^t e_t+\sum_{i=0}^{t-1}\alpha_i r^i e_i=\sum_{i=0}^t \Big\lfloor \frac{m}{r^i}\Big\rfloor,
      $$
where $e_t$ is defined in \eqref{dikla6} 
\end{lemma}

{\vspace*{.1in}{\em Proof } of the upper bound on the distance \eqref{eq:d2}:} 
We need to prove that the distance of an $(n,k)$ code with $t$ disjoint recovering sets of size $r$ satisfies the inequality 
  $$
  d\leq n-\sum_{i=0}^t\Big\lfloor\frac{k-1}{r^i}\Big\rfloor.
 $$
Let $G$ be the recovering graph of the code. 
We will use Lemma \ref{best lemma} several times for the graph $G$. 
Assume that we are allowed to color $k-1$ vertices and would like to color them in the way that 
guarantees a large expansion ratio with respect to their closure.
We begin by using Lemma \ref{best lemma} for the graph $G_1=G$. According to it, $G_1$ contains
a subset $S_1$ of vertices of size at most $r^t$ whose 
expansion ratio is at least $e_t$. Color the vertices in $S_1$ and $\C(S_1).$ 
Then call $G_2$ the subgraph induced on the subset of vertices $V\backslash \C(S_1)$ and 
apply Lemma \ref{best lemma} to $G_2,$ etc. Continuing this process, suppose that in the $i$-th 
round there are $b_i$ vertices still to be colored, and let $G_i$ be the induced subgraph of $G$ 
on the set of vertices that have not been colored in the previous $i-1$ rounds.  
Each vertex in $G_i$ has outgoing edges of all $t$ colors because if not, then one of its recovering
sets has been already removed, but then this vertex itself cannot be present because of the definition
of the closure. 
Let $m\leq t$ be the largest integer such that $r^m\leq b_i$. 
Now apply Lemma \ref{best lemma} for the graph $G_i$ to find a set $S_i$ of vertices of size at most 
\begin{equation}
|S_i|\leq r^m
\label{dikla4}
\end{equation} 
and expansion ratio at least $e_m$. Now color the set $S_i$. 
Continue this process until we have used all the $k-1$ vertices and call the obtained set of $k-1$
vertices $S.$ In each step the cardinality of $S_i$ is at most $r^m$ according to
\eqref{dikla4}, and hence 
\begin{align}
\label{qwqw}
|\C(S)|\geq \lfloor \frac{k-1}{r^t} \rfloor r^t e_t + \sum_{i=0}^{t-1}\alpha_ir^ie_i,
\end{align}
where $$k-1=\sum_{i}\alpha_ir^i,$$
is the $r-$ary representation of $k-1$. Using Lemma \ref{best lemma2}, \eqref{qwqw} becomes 
$$|\C(S)|\geq \sum_{i=0}^t\Big\lfloor\frac{k-1}{r^i}\Big\rfloor.$$
Since the value of the coordinates in $\C(S)$ is determined by the value of the coordinates in $S$ which is of size $k-1$, the size of the restriction of the code $\cC$ to coordiantes $I=\C(S)$ is at most 
$|\cC_I|\leq q^{k-1}<q^k,$
hence 
$$d\leq n-|I|=n-|\C(S)|\leq n-\sum_{i=0}^t\Big\lfloor\frac{k-1}{r^i}\Big\rfloor.\hspace*{.5in}\qed$$
%Note that for codes with no locality, and codes with a single set the bound on the distance is tight. 
%distance is at most
%\begin{align*}
%d&\leq n-\lfloor \frac{k-1}{r}\rfloor r\cdot e_1- ((k-1)\mod r)\cdot e_0\\
%& =n-\lfloor \frac{k-1}{r}\rfloor (r+1)- (k-1)\mod r\\
%&=n-(k-1 +\lfloor \frac{k-1}{r}\rfloor),
%\end{align*}
%which is known to be a tight bound derived previously.

\vspace*{.05in}
{\em Proof} of Lemma \ref{best lemma}:
We apply induction on $t$. For $t=0$ there are no edges in the graph. Define $S=\{v\}$ and note that $\C(S)=S=\{v\},$ 
and the expansion ratio is $1$ as needed. Now assume that the claim is correct for $t$ recovering sets. Let us prove it for 
$t+1$ recovering sets. Remove from $G$ the vertex $v$. For each other vertex $u\neq v$ we remove the edges that correspond 
to one of its recovering sets. Specifically, if $u$ has a recovering set that contains $v$, we
remove all of its edges that correspond to this recovering set; otherwise, remove the edges that correspond to any one of its recovering sets. 
Denote the resulting graph by $G_1$, and observe that each vertex of $G_1$ has exactly $t$ recovering sets. 

Let $v_1,...,v_l$ be the vertices of one of the recovering sets of $v$, where $l\leq r$. Our plan is to apply the induction hypothesis successively $l$ times for some induced subgraphs of $G_1$ which we denote below by $G_i, i=1,\dots, l$. 
We also use the notation $\C_i(S), i=1,...,l$ to refer to the closure operation of the set $S$ in the graph $G_i$, 
and use the notation $\C(S)$ to refer to the closure operation in the original graph $G$. Upon performing the $i$th step we
will have the vertex $v_i$ colored.

In the first step, we use the induction hypothesis to find a set $S_1$ of size at most $r^t$ in the graph $G_1$ whose expansion ratio 
is at least $e_t,$ and such that $v_1\in \C_1(S_1)$. 
Suppose that $S_1,\dots,S_{i-1}$ sets of vertices have been constructed in the first $i-1$ steps, $2\le i\le l.$ Denote by 
$G_i$ the graph $G_1$ obtained upon removing the set of vertices $\C_1(S_1\cup ...\cup S_{i-1}).$ 

Let us describe the construction of the set $S_i$.
If $v_i \in \C_1(S_1\cup ...\cup S_{i-1}),$ put $S_i=\emptyset.$ Otherwise $v_i\in V(G_i)$. 
Note that each vertex $u$ in $G_i$ has outgoing edges of all $t$ colors because otherwise, if $u$ is missing one color, then it has a recovering set that is contained in $\C_1(S_1\cup ...\cup S_{i-1})$, and then also $u\in \C_1(S_1\cup ...\cup S_{i-1}).$ 
Apply the induction hypothesis for $G_i$ to find a set $S_i$ of size at most $r^t$ and expansion ratio at least $e_t$ 
such that $v_i\in \C_i(S_i)$. Notice that since $\C_i(S_i)$ is a subset of the vertices of the graph $G_i$, it is disjoint from the set $\C_1(S_1\cup...\cup S_{i-1})$. We claim that 
   $$
   S=\cup_{i=1}^lS_i
   $$ 
   is the desired set.
Observe that
\begin{align*}
\C_1(S)&=\C_1(S_1\cup...\cup S_l)\nonumber\\
&=\cup_{i=1}^l\C_1(S_1\cup...\cup S_i)\backslash \C_1(S_1\cup...\cup S_{i-1}).\nonumber\\
\end{align*}
Since 
  $$
  \C_1(S_1,\dots,S_i)=\C_1(S_1,\dots,S_{i-1})\cup \C_i(S_i)
  $$
(disjoint union), we obtain
   \begin{equation}
   \C_1(S)=\cup_{i=1}^l\C_i(S_i)\label{dikla},
   \end{equation}
where the union is also disjoint.

We claim that for any $i=1,...,l$ the vertex $v_i$ belongs to $\C_1(S)$. 
Indeed, by construction, if $S_i$ is the empty set, then $v_i\in \C_1(S_1\cup...\cup S_{i-1})$, otherwise $v_i\in \C_i(S_i)$. 
We conclude that $\C_1(S)$ contains a complete recovering set $v_1,...,v_l$ of the vertex $v$, and therefore, 
\begin{equation}
\C(S)=\C_1(S)\cup \{v\}.
\label{dikla2}
\end{equation}
%Hence $v_i\in C(S)$, and then also $v\in C(S)$, since $C(S)$ contains the recovering set $v_1,...,v_l$. 
The size of $S$ satisfies
  $$|S|=|\cup_{i=1}^lS_i|= \sum_{i=1}^l|S_i|\leq r\cdot r^t= r^{t+1},$$
and all is left to show is the expansion ratio. By \eqref{dikla} and \eqref{dikla2} 
\begin{align*}
|\C(S)|&=|\cup_{i=1}^l\C_i(S_i)\cup \{v\}|=1+\sum_{i=1}^l|\C_i(S_i)|.
\end{align*}
Hence the expansion ratio of the set $S$ satisfies  
\begin{align}
\frac{|\C(S)|}{|S|}&=\frac{1+\sum_{i=1}^l|\C_i(S_i)|}{|S|}\nonumber\\
& \geq \frac{1}{r^{t+1}}+ \frac{\sum_{i=1}^l|\C_i(S_i)|}{|S|}\nonumber\\
&= \frac{1}{r^{t+1}}+ \sum_{i=1}^l \frac{|S_i|}{|S|}\frac{|\C_i(S_i)|}{|S_i|}\nonumber\\
&\geq \frac{1}{r^{t+1}}+ \sum_{i=1}^l \frac{|S_i|}{|S|}e_t \label{dikla3}\\
&=\frac{1}{r^{t+1}}+ e_t\nonumber\\
&=e_{t+1},
\end{align}	 
where \eqref{dikla3} follows since the set $S_i$ has expansion ratio of at least $e_t$ in $G_i$. \hfill\qed

\vspace*{.05in}
{\em Proof} of Lemma \ref{best lemma2}:
%$$\lfloor \frac{k-1}{r^t}\rfloor r^t e_t+\sum_{i=0}^{t-1}\alpha_i r^i e_i=\sum_{i=0}^t \lfloor \frac{k-1}{r^i}\rfloor.$$
We apply induction on $t$. For $t=0$ the equality can be easily checked. We assume correctness for $t$ and prove it for $t+1$.
\begin{align}
&\Big\lfloor \frac{m}{r^{t+1}}\Big\rfloor r^{t+1} e_{t+1}+\sum_{i=0}^{t}\alpha_i r^i e_i\nonumber\\
&= \sum_{i=0}^{t+1}\Big\lfloor \frac{m}{r^{t+1}}\Big\rfloor r^{i}+\sum_{i=0}^{t}\alpha_i r^i e_i \nonumber
\end{align}
\begin{align}
&=\Big\lfloor \frac{m}{r^{t+1}}\Big\rfloor+\sum_{i=1}^{t+1}\Big\lfloor \frac{m}{r^{t+1}}\Big\rfloor r^{i}+\sum_{i=0}^{t}\alpha_i r^i e_i\nonumber\\
&=\Big\lfloor \frac{m}{r^{t+1}}\Big\rfloor+\sum_{i=1}^{t+1}\Big\lfloor \frac{m}{r^{t+1}}\Big\rfloor r^{i}+\alpha_t r^t e_t+\sum_{i=0}^{t-1}\alpha_i r^i e_i\nonumber\\
&=\Big\lfloor \frac{m}{r^{t+1}}\Big\rfloor+\sum_{i=1}^{t+1}\Big\lfloor \frac{m}{r^{t+1}}\Big\rfloor r^{i}+\alpha_t (\sum_{i=0}^tr^i)+\sum_{i=0}^{t-1}\alpha_i r^i e_i\nonumber\\
&=\Big\lfloor \frac{m}{r^{t+1}}\Big\rfloor+\sum_{i=0}^tr^i\Big(\Big\lfloor \frac{m}{r^{t+1}}\Big\rfloor r+\alpha_t\Big)+\sum_{i=0}^{t-1}\alpha_i r^i e_i\nonumber\\
&=\Big\lfloor \frac{m}{r^{t+1}}\Big\rfloor+\sum_{i=0}^tr^i\Big\lfloor \frac{m}{r^{t}}\Big\rfloor+\sum_{i=0}^{t-1}\alpha_i r^i e_i\nonumber\\
&=\Big\lfloor \frac{m}{r^{t+1}}\Big\rfloor+\Big\lfloor \frac{m}{r^{t}}\Big\rfloor r^te_t+\sum_{i=0}^{t-1}\alpha_i r^i e_i\nonumber\\
&=\Big\lfloor \frac{m}{r^{t+1}}\Big\rfloor+\sum_{i=0}^t\Big \lfloor \frac{m}{r^i}\Big\rfloor \label{dikla5}\\
&=\sum_{i=0}^{t+1} \Big\lfloor \frac{m}{r^i}\Big\rfloor \nonumber,
\end{align}
where \eqref{dikla5} follows from the induction hypothesis, and the result follows.

\providecommand{\bysame}{\leavevmode\hbox to3em{\hrulefill}\thinspace}
\providecommand{\MR}{\relax\ifhmode\unskip\space\fi MR }
% \MRhref is called by the amsart/book/proc definition of \MR.
\providecommand{\MRhref}[2]{%
  \href{http://www.ams.org/mathscinet-getitem?mr=#1}{#2}
}
\providecommand{\href}[2]{#2}

%\bibliographystyle{IEEEtran}
%\bibliography{polar}

\end{document}